\documentclass[12pt]{article}
\usepackage{graphicx,amsmath}
\usepackage{graphicx,amsmath,lineno,color}

\usepackage{units}

\newlength{\dinwidth}
\newlength{\dinmargin}
\renewcommand{\vec}[1]{\boldsymbol{#1}}

\def\lapproxeq{\lower .7ex\hbox{$\;\stackrel{\textstyle                                                    
<}{\sim}\;$}}                                                    
\def\gapproxeq{\lower .7ex\hbox{$\;\stackrel{\textstyle                                                    
>}{\sim}\;$}}                                                    
\def\be{\begin{equation}}                                                    
\def\ee{\end{equation}}                                                    
\def\bea{\begin{eqnarray}}                                                    
\def\eea{\end{eqnarray}}

\def\sh{\hat s}
\def\sh2{{\hat s}^2}

\begin{document}

\titlepage                                                    
\begin{flushright}                                                    
\today \\                                                    
\end{flushright} 
\vspace*{0.5cm}

\begin{center}                                                    
{\Large \bf  Bethe phase variation due to a non-exponential nuclear amplitude and the possibility of  using a  $t$-dependent phase to determine the  $\rho$-parameter from elastic scattering data}\\

\vspace*{1cm}

P. Grafstrom$^{a}$ and M.G. Ryskin$^{b}$ \\                                                   
                                                   
\vspace*{0.5cm}                                                    
$^a$ Universit\`{a} di Bologna, Dipartimento di Fisica , 40126 Bologna, Italy \\
$^b$ Petersburg Nuclear Physics Institute, NRC Kurchatov Institute, Gatchina, St.~Petersburg, 188300, Russia

\vspace*{1cm}

\begin{abstract}
We evaluate the possible deviation (from the conventional Cahn's result) of the phase between the one-photon-exchange and the `nuclear' high energy 
$pp$ scattering amplitudes in a small $t\to 0$ region caused by a more complicated (not just $exp(Bt)$)  behaviour of the nuclear amplitude. Furthermore we look at the possible role of the {\em $t$-dependence} of the  $\rho(t) \equiv$ Real/Imaginary amplitude ratio. It turns out that both effects are  rather small - much smaller than to have any influence on the experimental accuracy of  $\rho(t=0)$ extracted from the elastic proton-proton scattering data.
\end{abstract}

\end{center}

\vspace{1cm}

\section{Introduction}
The real part of the high energy strong interaction (nuclear) $pp$ elastic amplitude, $F^N$, was measured recently by TOTEM collaboration at $\sqrt s=13$ TeV with unprecedented accuracy of 0.01 for the $\rho=$Re$F^N(t=0)$/Im$F^N(t=0)$ ratio~\cite{TOTEM}.\\
The conventional way to measure the real part of the strong interaction (nuclear)
forward amplitude is to consider its interference~\footnote{it is called the Coulomb Nuclear Interference (CNI).} with the pure real one-photon-exchange QED amplitude, $F^C$, at very small momentum transfer $t\to 0$. However this interference is affected by the possibility of  multiphoton exchange processes which 
result in the additional phase difference $\alpha\Phi$.
That is the  total amplitude reads
\begin{equation}
\label{1}
F^{TOT}~=~F^N~+~e^{i\alpha\Phi}F^C\ .
\end{equation}
Here $\alpha=\alpha^{\rm QED}=1/137$. The phase $\Phi$ (the so-called Bethe phase) was calculated first by Bethe~\cite{Be} using
 the WKB approach, and then was re-examined by West and Yennie \cite{WY} in terms of Feynman diagrams. A more precise calculation was performed by R. Cahn \cite{Cahn} in 1982 which accounts for the details of the proton form factor. It gives
\begin{equation}
\label{2}
\Phi(t)~=~-[\ln(-Bt/2)+\gamma_E+C]\ ,
\end{equation}
where $B$ is the $t$-slope of the elastic cross section
($d\sigma_{\rm el}/dt\propto e^{Bt})\ $, $\gamma_E=0.577...$ is Euler's constant and the constant 
$C\sim 0.4 -0.6$   depends on the precise form of the proton
 electromagnetic form factor and the $t$ dependence of the nuclear amplitude. In the Cahn's paper~\cite{Cahn} the usual dipole electromagnetic  formfactor $f(t)=1/(1-t/0.71\mbox{GeV}^2)^2$ was used and the pure exponential $t$ dependence of $F^N\propto\exp(Bt/2)$ was assumed. In such a case the value of $C=0.60$ for CERN-ISR energies and $C=0.45$ for the LHC case when the slope $B\simeq 20$ GeV$^{-2}$.\\
 
 However the real $t$ dependence of nuclear amplitude is more complicated even at a rather low $|t|$. In particular the deviation from the pure exponent was observed at 8 TeV in ~\cite{Tot1,Tot2}. This deviation should affect the constant $C$ calculation and the main aim of the present paper is to evaluate how large this effect  can be.
We also evaluate in section 3  the effect of a $t$-dependence of the nuclear phase on the determination of the $\rho$-parameter   from elastic scattering at low $|t|$.
\section{Estimate of the constant $C$ alteration} 
Note that thanks to a small QED coupling $\alpha=1/137$ the absolute value of the phase $\alpha \Phi$ is small. Within the relevant for Coulomb-nuclear interference interval $|t|=0.001 - 0.03$ GeV$^2$ it does not exceed 0.03. That is actually the reason why we are looking for the contribution of the first diagram with one additional photon exchange (see Fig.1).

 \begin{figure}[h]
 \vspace{-5cm}
 \begin{center}
\includegraphics[scale=0.5]{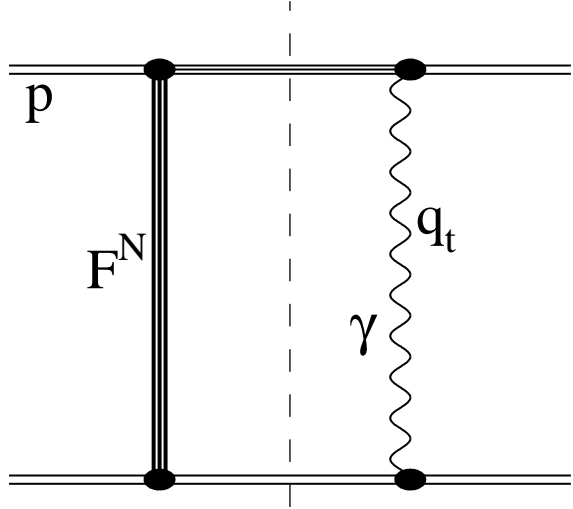}
\caption{\sf Diagrams responsible for the Bethe phase at first $\alpha^{\rm QED}$  order. The nuclear amplitude is shown by the triple  solid  line and marked as $F^N$}
\end{center}
\label{fp1}
\end{figure}
Next term in the $e^{i\alpha\Phi}$ expansion, $(\alpha\Phi)^2/2$,  is of the order of 10$^{-3}$ and the  small possible change of this contribution is already negligible in comparison with the present experimental accuracy $\sim 10^{-2}$. Moreover, actually this second term interferes
only with the real part of $F^N$. That is the expected effect should be
 of about $\rho(\alpha\Phi)^2/2\lapproxeq 10^{-4}$~\footnote{Here in the intermediate state we consider only the proton. The possibility of the $p\to N^*$ excitation was studied in~\cite{KMR-B}. It was shown that the effect of these possible additional contributions on the value of $\rho$  does not exceed $10^{-3}$.}\\
 
 Thus to estimate the $C$ variation caused by a more complicate $t$ dependence of the nuclear amplitude we have to compare the contributions of  Fig.1 diagram with the complete $t$ dependence of $F^N_{exact}$ with that with $F^N_{exp}(t)=F^N(0)\exp(Bt/2)$.
 
 \begin{equation}
 \label{e1}
 \delta C(t)~=~-\int d^2q_t\frac{f^2(q^2_t)}{q^2_t}\left(\frac{F^N_{exact}(t')}{F^N_{exact}(t)}  -\frac{F^N_{exp}(t')}{F^N_{exp}(t)} \right)\ .
 \end{equation}
We denote the full transverse momentum transferred as $Q_t$ ($t=-Q^2_t)$. So the momentum transferred through the nuclear amplitude in Fig.1 is $\vec Q'_t=\vec Q_t-\vec q_t$ and $t'=-Q^{'2}_t$.

Note that the integral (\ref{e1}) has no infrared divergence. The difference $[F_{exact}(t')/F_{exact}(t)-F_{exp}(t')/F_{exp}(t)]\to 0$ at  $q_t\to 0$ since $t'\to t$. 
 
 The integral (\ref{e1}) was computed numerically for the case of 8 TeV $pp$ scattering using the TOTEM parameterization of the  nuclear amplitude $F^N_{exact}(t)=F^N_{exact}(0)\exp(-\frac 12\sum_1^3 b_k|t|^k)$. The slope of $F^N_{exp}$ amplitude was taken to be $B=B(t=0)=b_1$ and the  electromagnetic formfactor has the dipole form $f(q^2_t)=1/(1+q^2_t/0.71\mbox{GeV}^2)^2$ .
 
\begin{figure}[h]
  \vspace{-4cm}
  \begin{center}
\includegraphics[scale=0.5]{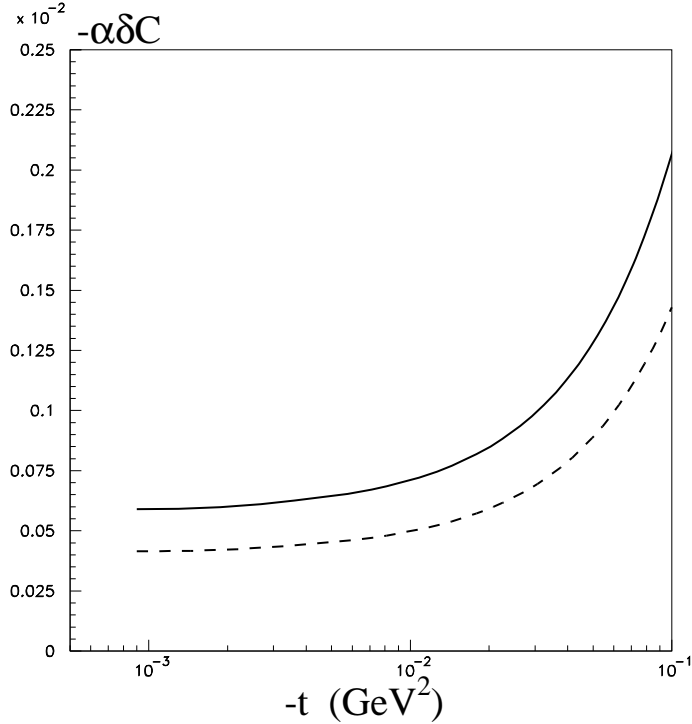}
\caption{\sf The deviation of the phase between the one photon exchange and the nuclear, $F^N(t)$, amplitudes caused by the more complicated $t$ dependence of $F^N_{exact}$ in comparison with the pure exponential behaviour used in Cahn's~\cite{Cahn} calculations. Continues curve corresponds to the parameterization of~\cite{Tot2}, which accounts for the Coulomb-nuclear interference, while for the dashed curve the parameters of~\cite{Tot1} (without the Coulomb term) was used. }
\end{center}
\label{p2}
\end{figure}

  The results are shown in Fig.\ref{p2}. Continues curve corresponds to the parameterization of ~\cite{Tot2}
 which accounts for the Coulomb-nuclear interference while for the dashed curves the parameterization of~\cite{Tot1}, based on the description of a larger $|t|$ interval (without the Coulomb term) was used.

It is seen from Fig.\ref{p2} that the possible phase shift $\alpha\delta C$ never exceed 10$^{-3}$ in the $|t|<0.03$ GeV$^2$ region relevant for the Coulomb-nuclear interference.\\
 This is consistent with the very naive estimate. Since the difference between the exact $pp$ nuclear amplitude and its exponential approximation at low $|t|$ is less than 10\% we can expect $\delta C<0.1$; that is the phase shift $\alpha \delta C<10^{-3}$. One can safely  neglect this effect and use the Cahn's expression~\cite{Cahn}, written for the pure exponential $F^N(t)\propto \exp(Bt/2)$ case (with $B=B(t=0)$), bearing in mind the experimental accuracy of the order of 10$^{-2}$.

\section{$t$-dependence of $\rho$}
Another point which should be considered is the following. Actually the value of $\rho(0)$ is extracted from the elastic scattering differential cross section $d\sigma_{el}/dt$ measured not at $t=0$ but in some interval of small but non-zero $t$. On the other hand we know that $\rho(t)\neq const(t)$. The real part of the elastic amplitude should vanish at some relatively low$|t|\sim 0.1$ GeV$^2$~\cite{Martin}   (see e.g. Fig.1 of~\cite{zero}  as an example).  The corresponding $t$ dependence of the Re/Im ratio may also affect the value of $\rho(0)$ obtained by fitting the $d\sigma/dt$ data under the usual/simplified assumption that $\rho(t)=const(t)$.

The most straightforward way to take into account the fact that the real part should vanish at $|t|\sim 0.1$ GeV$^2 $  is to use the following simplified formula for the $t$-dependence of the nuclear phase

\begin{equation}
\label{Misha}
\arg F^N(t)=\frac\pi 2-\arctan\left(\rho(0)(1+t/0.1)\right)
\end{equation}

 A similar $t$-dependence was also recently proposed by Durand-Ha \cite{Durand}  taking into account in addition the fact that the imaginary part  has a zero around  the dip region $|t|\sim 0.45$ GeV$^2$~.

\begin{equation}
\label{Dur}
\arg F^N(t)=\frac\pi 2-\arctan\left(\rho(0)\frac{1+t/t_R}{1+t/t_I}\right)
\end{equation}

with $t_R=0.16$ GeV$^2$ and $t_I=0.42$ GeV$^2$ at 13 TeV.

Naively one would expect a small effect here because looking at the corresponding $t$-dependence of
$\rho$ one sees a very weak dependence of $\rho$ in the coulomb interference region. At the point of
maximum sensitivity to the interference effect the deviation between the $\rho$ value for a constant
phase and the phase from (\ref{Misha},\ref{Dur}) is less than $10^{-3}$ ( remember that the best experimental uncertainty up to now is $10^{-2}$).

To study this effect/question more quantitatively we have analyzed the published TOTEM 13 TeV data~\cite{TOTEM} using the $t$-dependent phase of (\ref{Misha},\ref{Dur}). We also tried a couple of other versions of  possible $t$-dependent phases  published earlier: the so called  standard parameterization~\cite{standard}
\begin{equation}
\label{st}
\arg F^N(t)=\frac \pi 2-\arctan\rho(0)+\arctan\left(\frac{|t|-|t_0|}\tau\right)-\arctan\left(\frac{-|t_0|}\tau\right)
\end{equation}
with $t_0=-0.5$ GeV$^2$ and $\tau=0.1$ GeV$^2$,

and the Bailly parameterization ~\cite{Bailly}
\begin{equation}
\label{Bail}
\arg F^N(t)=\frac \pi 2-\arctan\frac{\rho(0)}{1-t/t_0}
\end{equation}
with $t_0=-0.53$ GeV$^2$

 Besides those we consider a more  extreme hypotheses 

 the so called  "peripheral" model~\cite{peripheral} where~
\begin{equation}
\label{periph}
\arg F^N(t)=\frac\pi 2-\arctan\rho(0)-\xi_1\left(-\frac t{1\mbox{GeV}^2}\right)^\kappa e^{\nu t}
\end{equation}
with $\xi_1=800$, $\kappa=2.311$ and $\nu=8.161$ GeV$^{-2}$ 
and for comparison just the 
\begin{equation}
\label{const}
\rho(t)=const
\end{equation}

\begin{figure}[h]
 \begin{center}
\includegraphics[scale=0.7]{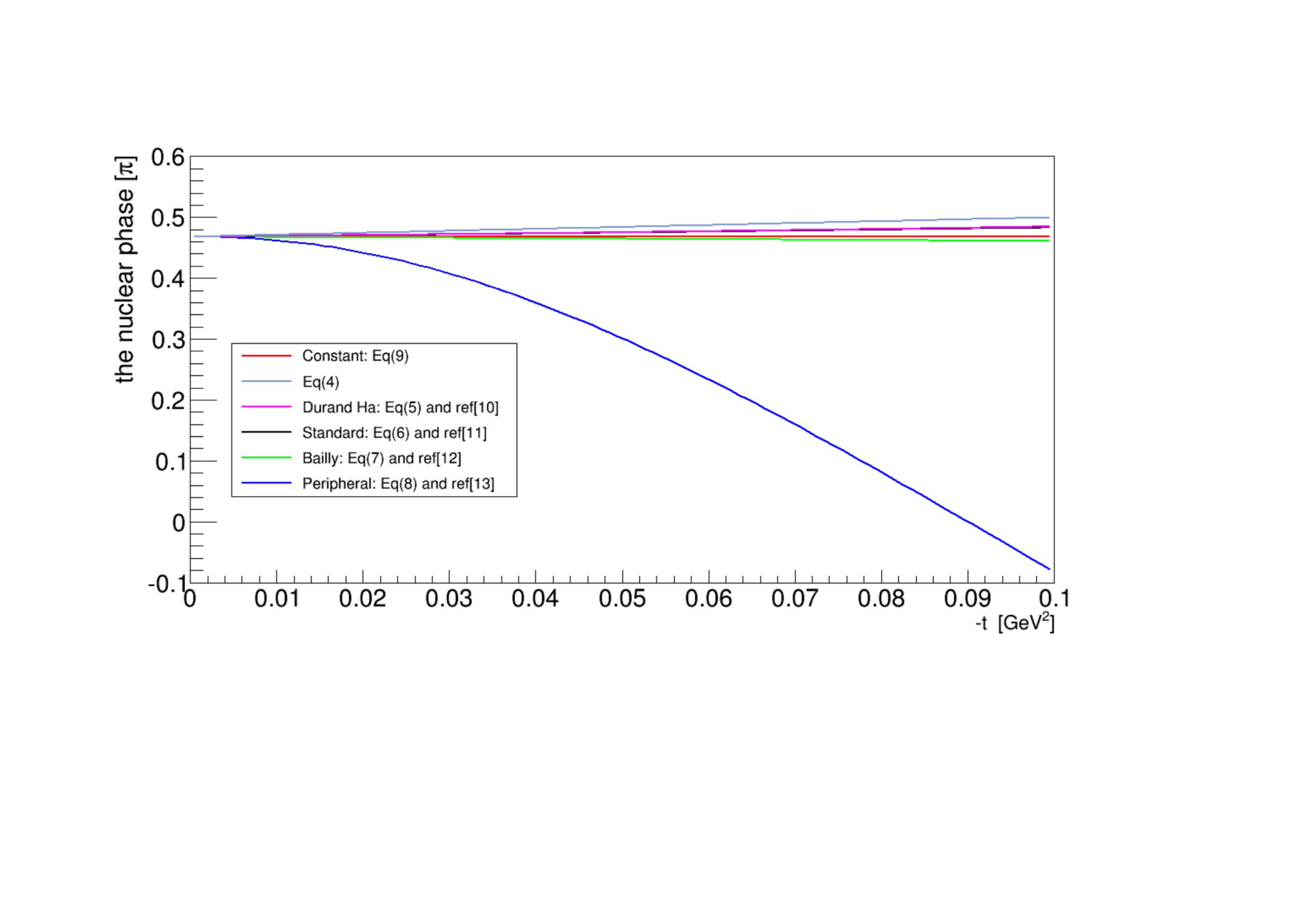}
\vspace{-4cm}
\caption{\sf $t$-dependence of the nuclear phase for the  six models~ref \cite{Durand,standard,Bailly,peripheral} and equation (4) and (9).}
\end{center}
\label{rho}
\end{figure}

The corresponding  $t$-dependence of all  the mentioned phases are shown in Fig.\ref{rho}.

In all these six cases the data~\cite{TOTEM} where fitted using the conventional Cahn's phase (\ref{2}). The nuclear amplitude was parametrized as
\begin{equation}
\label{BC}
|F^N(t)|^2=A\exp(Bt+Ct^2)\ .
\end{equation}

Thus we have four free parameters: $A,B,C$ and $\rho(0)$.

We have confirmed that, indeed, the difference in $\rho$ for the case of a constant phase and the phase of (\ref{Misha}-\ref{periph}) is
less than $10^{-3}$ {\bf $^{}$ }
by fitting the TOTEM data 13 TeV in the $t$-range $0.0008$  GeV$^2$- $0.12$ GeV$^2$and using the parameterization (\ref{BC}). The only exception is the "peripheral" model. In this model the value of $\rho(0)$ differs from that in the $\rho=const(t)$ case by about $6\cdot 10^{-3}$ but still smaller than the typical experimental uncertainty of $10^{-2}$.~\footnote{Since the parameters for the peripheral model/scenario at 13 TeV were not published we have used the numbers from 8 TeV  \cite{Tot2}. Due to the weak/logarithmic behaviour of elastic pp-amplitude the difference should not be large while on another hand all this examples are just to demonstrate the expected order of the  size of  the effect. The same $6\cdot 10^{-3}$ difference was  observed fitting  with this   $t$-dependence of $\rho$   the 8 TeV data \cite{Tot2}.\\ } \\
Note however that the peripheral model (\ref{periph}) is inconsistent with the dispersion relations for the C-even amplitude. In the simplified form the dispersion relation at fixed $t$ reads~\footnote{Here we use the 2Im$F^N(0)=\sigma_{tot}$ normalization.}
\begin{equation}
\label{disp}
\rho~\simeq~\frac\pi 2\frac{\partial\ln(\mbox{Im}F^N(t))}{\partial\ln s}\ . 
\end{equation} 
 
As it follows from the experimental data this value
should be positive in the $|t|\sim 0.1$ GeV$^2$ region while
in the peripheral model it becomes negative (see Fig.\ref{rho}). 

Thus we conclude that both effects - the possible $t$-dependence of the $\rho=$Re$F^N(t)$/Im$F^N(t)$   and some deviation from the pure exponential $t$ behaviour of the nuclear amplitude $F^N(t)$ in the small $|t|$ region relevant for extraction of $\rho(t=0)$, via the Coulomb-nuclear interference, do not exceed the 
10$^{-3}$ level and can be neglected in comparison with the today experimental accuracy of about 10$^{-2}$.  \\

{\bf\Large Acknowledgements}\\

We thanks V.A. Khoze for the reading of this manuscript.
\end{document}